# Resiliency by Retrograded Communication

The Revival of Shortwave as a Military Communication Channel

Jan Kallberg
Army Cyber Institute at West Point and the Department of Social Sciences
United States Military Academy
West Point, NY 10996
Jan.kallberg@westpoint.edu

Stephen S. Hamilton
Army Cyber Institute at West Point and the Department of Electrical Engineering and Computer Science
United States Military Academy
West Point, NY 10996
Stephen.hamilton@westpoint.edu

*Abstract – In the last three decades, the great powers have become increasingly dependent on satellite communication (SATCOM), very high frequency (VHF), and ultra-high frequency (UHF) providing high bandwidth line of sight (LOS) communications. These military communication channels lack resilience because an EW campaign can affect both VHF and SATCOM simultaneously. The 1940s preferred spectrum, high frequency (HF), with its different propagation patterns, offers an opportunity for military communication resiliency in the 21st century. The concept of retrograding could give an operational advantage and create the ability to sustain communication in electronic warfare (EW) saturated environment.*

## I. Introduction

Defense forces, coast guard, homeland security, and emergency management agencies have, during the last decades, become dependent and reliant on stable communications providing ample bandwidth to support operations and information flow within the organization. The current deployed tactical and operational communication networks are highly dependent upon line-of-sight (LOS) communication utilizing very high frequency (VHF), ultra-high frequency (UHF), and higher frequencies. The information grids are tailored by a network of radio, datastream, radio links, and satellite communication (SATCOM), all requiring a near to line-of-sight for sufficient propagation, bandwidth, and transmission quality.

Meanwhile, the efforts to disrupt and degrade military communications are equally focused on the VHF/UHF bands that are prevailing for tactical and operational radio and data traffic. Within the last decade, especially Russia and China, have made a dedicated effort to radically improve their ability in electronic warfare (EW) and create effects in the electromagnetic spectrum (EMS) in the pursuit to deny or degrade targeted communication channels. Parallel with this development SATCOM communication risks have increased as several states developed electromagnetic, kinetic, and cyber ability to disrupt these communications.

A potential future conflict with a capable near-peer adversary; Russia or PR China are notable in that they have heavily invested in electromagnetic spectrum warfare capacity, and are capable of employing electronic warfare throughout their force structure [1]. Electronic warfare elements deployed within theaters of operation threaten to degrade, disrupt, or deny VHF, UHF, and SATCOM communication.

In this scenario, HF radio is a viable backup mode of communication. The ability retrograde to HF, commonly referred to as shortwave, creates resilience in an EW-saturated operational environment. The HF networks provide satisfying bandwidth if the focus is to transmit operational information, command, and control.

## II. Historical Path

The first globally operating military force, with a presence on all continents, was the British Army and the Royal Navy. They were instrumental in the growth of the British colonial possessions. Once the British Empire was established, there was a need to communicate with British colonies and direct a British military presence. This paved the way for financing and incentives to rapidly create a network of undersea cables that linked the colonies with Great Britain in the 1850s and 1860s [2]. The telegraphy, as a form of domestic communication, had followed the unprecedented increase in transportation infrastructure in the early 1800s. The US railway system in 1830 was 40 miles, and by 1860, 920 miles of railroad been laid down [3]. The telegraph lines followed the railroads, and during the American Civil War, 1861-1865. Both sides used the telegraph to instruct and command units during the engagement.

In comparison to landline, the earlier undersea cables struggled with extremely low bandwidth down to a word a minute, which increased over time. The great powers of the time, the United Kingdom, the United States, France, Russia, and Germany had in the early 1900s established land-based networks of wireless telegraph stations. At the outbreak of World War One, the great powers of the day had outfitted their larger warships with wireless telegraphs using low frequency (LF), also called longwave, and medium frequency (MF), known as medium wave, providing communication 80 – 200 km between warships. Due to the size and electricity consumption of the early wireless telegraph, the military wireless equipment for armies had a limited range, and the early airborne wireless telegraphs had ranges that were less than 2 km.

The introduction of shortwave in the 1920s gave governments, companies, and individuals the ability to communicate over unprecedented distances at a low cost [4]. The innovation of more efficient transmitters also made it possible to design and produce mobile two-way radio [4]. During World War Two, the Korean War, and the War in Vietnam, HF was a standard way of communicating that was gradually taken over by VHF during the War in Vietnam. The successful launch of Telstar [5] in 1962 and the evolution of satellite communication during late 1960 changed the flow of communications [6]. In the early 1960s, HF and cable were utilized to communication to the US or higher commands in theater. Post-Vietnam and the early 1980s, tactical VHF, radio link, and SATCOM were utilized for communication to higher commands. From the 1980s, this development towards VHF/UHF LOS communications has accelerated. Since the early 1990s HF is rarely utilized by ground forces. Even if HF equipment is fielded it is seldom used. However
, HF continues to be used for limited tasks within navies and air forces.

### III. THE BANDWIDTH ADDICTION

The vast majority of the world's modern armies' ability to employ high-frequency HF radio systems has atrophied significantly since the Cold War as NATO, and numerous other countries transitioned to counterinsurgency operations. An era of abundant and close-to-uninterrupted bandwidth is coming to an end as the future threat landscape, and potential conflicts will likely involve aggressive EW, counter-SATCOM [7], and anti-satellite measures [8]. Also, satellites can be destroyed by cascading events due to debris from kinetic anti-satellite attacks on other spaceborne assets [9] or space debris already existing in orbit [10].

After two decades with the uncontested spectrum, the US Armed Forces, and a vast majority of the world's modern armies, have been used to having available bandwidth, communications, and the ability to switch between communication channels with limited interruption and excellent quality. The counterinsurgency operations have provided rear operational areas with a stable energy supply, enabling the ability to set up satellite and radio links, and reliable communication channels to higher commands, air assets, medical resources, and the logistics chain.

Any potential near-peer adversaries are fully aware of our dependence on these communications channels and how the loss of these channels would severely impact the modern way of warfighting. Satellite communications are especially vulnerable for several reasons. First, the satellites transmit at lower power levels, making them easier to jam. Second, weather and space weather (solar flares) can negatively impact satellite communications. Third, the compact and fragile design of the satellite itself makes them subject to failure due to space debris or potentially an attack from an adversary's satellite. Finally, the satellites can be challenging to upgrade and could, over time, be vulnerable to cyber-attacks.

William J. Lynn, III, the former US Deputy Secretary of Defense, stated in the summer of 2011 [11]: "*The willingness of states to interfere with satellites in orbit has serious implications for our national security. Space systems enable our modern way of war. They allow our warfighters to strike with precision, to navigate with accuracy, to communicate with certainty, and to see the battlefield with clarity. Without them, many of our most important military advantages evaporate.*"

Even if the US today is dominant in military satellite communication, several other countries are quickly building a spaceborne information infrastructure, where Russia, China, and India have taken the lead.

The knowledge that the high-bandwidth, almost always available, networks will likely be at least degraded in a future conflict should drive a cultural shift towards a more frugal usage of communication that reemphasizes the exchange to what is mission essential.

In the past, the radiotelegraph operators, and the organizations that utilized the Morse code, solved the lack of bandwidth by compressing messages. As an example, "2 GERSUB 10 NM SSW GIB" transmitted in a few words reads two German submarines are sighted ten nautical miles South-South-West of Gibraltar. During WW I and WW II trained radiotelegraphists were able to send and receive information that managed ground, air, and naval operations over vast spaces. The notion that current operations need megabits per second in available bandwidth is a sign that today's armies, coast guard, homeland security, and emergency management have grown accustomed to having access to high bandwidth and adjusted the processes to require these digital streams. The counterinsurgency battlefield, with bases and forward operating bases with a full array of information systems and communication infrastructure, close to the area of operation, was able to deliver high bandwidth close to the actual engagement. The HF networks give a reduced bandwidth for data transfer, even if it as limited as a 1990s Hayes 9600 baud dialup modem, but that would be enough if the exchange is compressed information that is mission essential. The Harris PRC-150 HF radio operates at 9600 bits per second using the STANAG 4539 NATO standard protocol.

### IV. REEMERGED OPERATIONAL ENVIRONMENTS

The recent change from a strategic focus on counterinsurgency to near-peer and peer-conflicts also extend the areas of operations. The counterinsurgency operational environment allowed the establishment of communication hubs in "safe zones" close to the battlefield. Even if these nations, where counterinsurgency operations occurred, lacked a national communications infrastructure, the military communication infrastructure could be established and sustain the operation. As General Milley stated [12], "Many of the conditions we have grown accustomed to over the past eighteen years will not exist in future battles. Control of the air will be contested; Forward Operating Bases will not provide a safe haven; units will be continuously targeted by enemy fires; and communications and navigation systems will be intermittent at best."

Apart from the Syrian civil war and the conflict in Eastern Ukraine, the conflicts the last two decades has had an uncontested spectrum and access to VHF/UHF with no adversarial EW assets present. These conditions will likely change, especially in a peer or near-peer conflict.

The future operations could occur in areas with limited satellite coverage to enable military satellite communication over a military space information grid or leased channels from a commercial operator using mobile satellite services (MSS). In that case, even with no present EW or interference, SATCOM can be unreliable and intermittent.

V. THE STATE OF HF (SHORTWAVE) COMMUNICATIONS

Currently, the competency with HF radio systems within the US Army is limited; however, there is a strong case to train and ensure readiness for the utilization of HF communication. Even in EMS-denied environments, HF radios can provide stable, beyond-line-of-sight communication permitting the ability to initiate a prompt global strike. While HF radio equipment is also vulnerable to electronic attack, it can be difficult to target due to near vertical incident skywave signal propagation. This propagation method provides the ability to reflect signals off the ionosphere in an EMS-contested environment, establishing communications beyond the line of sight. Due to the signal path, the ability to target an HF transmitter is much more complicated than transmissions from VHF and UHF radios that transmit line of sight ground waves.

One concern is the HF capacity, once seen as obsolete and replaced by VHF/UHF, has been removed to free up space and lower weight in several fixed-wing, helicopter, and vehicle assets. In some cases, versions of a particular platform can differ in the ability to communicate using HF. Where the older version had the HF capable radios delivered from the factory in the 1990s; meanwhile, the updated version had HF radios removed. The removal of HF equipment requires retrofitting HF ability back into the platform. There are needs to retrograde by adding, modifying, and updating the HF capacity in these platforms. Even if the equipment is fielded to the fighting formations, the ability across the branches will be fragmented and not uniform unless a DOD plan is in place to ensure their compatibility.

VI. RETROGRADE FOR RESILIENCE

The Russian investment in electronic warfare capabilities are significant, and electronic warfare units are organic to any Russian formation from the brigade combat team and higher. This can provide a significant strategic advantage in the early stage of a conflict. The Russian formations can engage cyber and electromagnetic effects already in the "Initial Period of War," within the grey zone before there is an open conflict [13].

The US and allied ground forces can offset initial strategic inferiority with airpower, naval power, and global strike abilities, but this is dependent on communication channels between ground forces and joint assets [13]. The focus of the adversary's electronic warfare is to deny US communications. One alternative is to retrograde and utilize high-frequency communications, which was the communication channel of World War II and the Korean War. High-frequency radio waves propagate by bouncing off the ionosphere allowing for beyond the line of sight communications. Due to the skywave propagation pattern, it is more difficult for the enemy to perform spectrum denial. Modern HF equipment is approximately the same in weight and size compared to VHF equipment and does not hinder usage by dismounted troops [14].

Even in EMS (electromagnetic spectrum)-denied environments, HF radios can provide stable, beyond-line-of-sight communication permitting the ability to initiate a prompt global strike. While HF (high frequency) radio equipment is also vulnerable to electronic attack, it can be difficult to target due to near vertical incident skywave signal propagation. This propagation method provides the ability to reflect signals off the ionosphere in an EMS-contested environment, establishing communications beyond the line of sight within a regional area. Due to the signal path, the ability to target an HF transmitter is more complicated than transmissions from VHF and UHF radios that transmit line of sight ground waves. Also, modern digital transmission modes allow for communications to occur at low power levels making it more difficult for the adversary to detect these signals.

Alarmingly, as hostile near-peer adversaries reemerge, it is necessary to re-establish HF alternatives should very-high-frequency, ultra-high frequency, or SATCOM come under attack and no longer remain viable options for battlefield communications. HF communication has its inherent weaknesses and challenges. Common challenges are antenna size and configuration and propagation changes throughout the day and night due to changes in the ionosphere. Still, it does not remove the fact that it can provide communications beyond the line of sight, which can serve as an alternative in critical junctures. By stepping back and being able to retrograde to HF as a resiliency measure, the US is increasing communication redundancy. This also adds an asymmetric advantage when the adversary has to divert EW assets with a different set of requirements to address the HF ability.

HF radio signals propagate by bouncing off the ionosphere and require more resources to disrupt and degrade. The HF propagation patterns would send signals to broader areas, which allows for the adversary to hear the signal and direct countermeasures. Still, it also will enable parts of the propagation to pass through enough to get communication established even in a high saturated EW environment. HF jamming equipment requires more energy and has a significant signature, which enables US and NATO neutralizing attacks with stand-off weaponry and anti-radiation missiles to be successful. The Russian Armed Forces utilize HF communications as well, and a broad and unrestricted HF jamming can degrade and disrupt their own communications. There is also a possibility that the HF transmission propagates in a way it cannot be heard by the adversary providing undisrupted communication when utilizing low probability of intercept (LPI) and low probability of detection (LPD) techniques. On the other hand, line-of-sight communications have a more narrow propagation channel, which allows the EW attacker higher certainty that communications are denied or degraded.

There is limited competency with HF radio systems within all the branches; however, there is a strong case to train and ensure readiness for the utilization of HF communication. Even in EMS (electromagnetic spectrum) denied environments, HF radios can provide stable, beyond-line-of-sight communication permitting the ability to initiate a prompt global strike. While HF radio equipment is also vulnerable to electronic attack, it can be difficult to target when configured to use near vertical incident skywave signal propagation. This high-angle take-off propagation method provides the ability to refract signals off the ionosphere in an EMS-contested environment, establishing communications beyond the line of sight out to 400 miles. Due to the high-angle signal path, the ability to direction find (DF) and target an HF transmitter is more complicated than transmissions from VHF and UHF radios that transmit line-of-sight ground waves. Also, Russian listening posts located outside of the 400-mile radius cannot intercept the communications. The recent digital modes utilizing 3G Automatic Link Establishment (ALE) technology allows for digital communication at lower power levels than what was previously required for voice. In addition, this mode simplifies operation by automatically selecting the optimal frequency for communication to other HF nodes in the network. This technology allows for tac chat messaging along with digital voice within a 3G ALE network. Using lower power is a crucial advantage when trying to prevent direction finding, and adding encryption to the digital signal helps prevent signal interception.

The HF spectrum offers low-cost opportunities to increase unit survivability and increase battlefield effectiveness by achieving a stealthier communication channel that potential adversaries will have difficulties to find the source of the transmission. HF should be included in any unit's communication PACE (Primary Alternate Contingency Emergency) plan as the emergency method at a minimum. The expense to attain an improved HF-readiness level is low compared to other defense initiatives, yet with a high return on investment. In the US, the equipment (Harris AN/PRC-150) has already been fielded to maneuver units, and foreign forces have similar equipment already purchased, which enables these forces to be retrograde to generate resiliency.

After almost three decades of limited interest in ground forces' HF communications, there are knowledge gaps to fill to ensure the optimal techniques, tactics, and procedures. Science and technology have during these decades advanced, therefore there multiple opportunities to cost-effectively enhance and improve the HF communication ability, especially pushing targeting data through HF communications. The revival of HF communications as a resilience measure will posture the Joint Force, and its branches, in a state of higher readiness for future conflicts.

## VII. CONCLUSION

In the last three decades, the great powers have become increasingly dependent on satellite communication (SATCOM) [15], very high frequency (VHF), and ultra-high frequency (UHF), providing high bandwidth line of sight (LOS) communications. These military communication channels lack resilience because an EW campaign can affect both VHF and SATCOM simultaneously. The 1940s preferred spectrum, HF, with its different propagation pattern, offers an opportunity for military communication resiliency in the 21$^{st}$ century.

Acknowledgment

We acknowledge the generous support of insights, technical and operational knowledge of HF, shared by Major Matthew G. Sherburne, U.S. Army, 1st Cyber Battalion, Cyber Protection Brigade, Fort Gordon, Georgia.

Authors

Jan Kallberg, Ph.D., is a research scientist at the Army Cyber Institute at West Point and an assistant professor at the US Military Academy. Col. Stephen S. Hamilton, Ph.D., is the chief of staff of the Army Cyber Institute at West Point and an academy professor at the US Military Academy.